\pdfoutput=1

\documentclass[preprint2, times, tighten]{aastex}

\usepackage{amsmath,amstext}
\usepackage[all]{hypcap} 

\usepackage{textcomp}
\usepackage{booktabs}
\usepackage{afterpage}
\usepackage{placeins}

\shorttitle{Upper Mass Limit for a Black Hole in the LMC}
\shortauthors{Boyce et al.}

\begin{document}

\title{An upper limit on the mass of a central black hole in the Large Magellanic Cloud from the stellar rotation field}
\author{H. Boyce\altaffilmark{1,2}, N. L\"utzgendorf\altaffilmark{2}, R. P. van der Marel\altaffilmark{2,3}, H. Baumgardt\altaffilmark{4}, M. Kissler-Patig\altaffilmark{5}, N. Neumayer\altaffilmark{6}, P. T. de Zeeuw\altaffilmark{7,8}}
\email{boyceh@physics.mcgill.ca}
\altaffiltext{1}{Department of Physics and McGill Space Institute, McGill University, 3600 University St., Montreal QC, H3A 2T8, Canada}
\altaffiltext{2}{Space Telescope Science Institute, 3700 San Martin Drive, Baltimore, MD, 21218, USA}
\altaffiltext{3}{Center for Astrophysical Sciences, Department of Physics \& Astronomy, Johns Hopkins University, Baltimore, MD 21218, USA}
\altaffiltext{4}{School of Mathematics and Physics, The University of Queensland, St. Lucia, QLD 4072, Australia}
\altaffiltext{5}{Gemini Observatory, Northern Operations Center, 670 N. A'ohoku Place, Hilo, Hawaii, 96720, USA}
\altaffiltext{6}{Max-Planck-Institut f\"ur Astronomie, K\"onigsstuhl 17, D-69117 Heidelberg, Germany}
\altaffiltext{7}{European Southern Observatory (ESO), Karl-Schwarzschild-Strasse 2, 85748 Garching, Germany}
\altaffiltext{8}{Sterrewacht Leiden, Leiden University, Postbus 9513, 2300 RA Leiden, The Netherlands}

\begin{abstract}
We constrain the possible presence of a central black hole (BH) in the
center of the Large Magellanic Cloud (LMC). This requires
spectroscopic measurements over an area of order a square degree, due
to the poorly known position of the kinematic center. Such
measurements are now possible with the impressive field of view of the
Multi Unit Spectroscopic Explorer (MUSE) on the ESO Very Large
Telescope.  We used the Calcium Triplet ($\sim850$nm) spectral lines
in many short-exposure MUSE pointings to create a two-dimensional
integrated-light line-of-sight velocity map from the $\sim 10^8$ individual spectra, taking care to identify
and remove Galactic foreground populations. The data reveal a clear
velocity gradient at an unprecedented spatial resolution of 1
arcmin$^{2}$. We fit kinematic models to arrive at a $3\sigma$
upper-mass-limit of $10^{7.1}$ M$_{\sun}$ for any central
BH - consistent with the known scaling relations for supermassive black holes and their host systems. This adds to the growing body of knowledge on the presence of BHs
in low-mass and dwarf galaxies, and their scaling relations with
host-galaxy properties, which can shed light on theories of BH growth
and host system interaction.
\end{abstract}

\keywords{black hole physics -- galaxies: individual (Large Magellanic Cloud) -- galaxies: kinematics and dynamics}
\maketitle

\section{Introduction}
As one of our closest neighbors, the study of the Large Magellanic Cloud (LMC) provides insights into many branches of astrophysics. These topics include studies of stellar populations \cite[e.g.][]{2000ApJ...542..804N,2013A&A...560A..44V}, the interstellar medium (ISM) \citep[e.g.][]{1990ARA&A..28..215D,2016AJ....151..161S}, microlensing by dark objects \citep[e.g.][]{2000ApJ...542..281A}, and the cosmological distance scale \citep[e.g.][]{2005ESASP.576..703N}. In addition, many recent photometric and kinematic datasets have shown that the inner regions of the LMC are dynamically complex \citep[e.g.][]{2000ApJ...545L..35Z,2002astro.ph..7077O}. An understanding of the structure and kinematics of the LMC is necessary for all of these applications. As a potential host of an intermediate-mass black hole (IMBH) or a supermassive black hole (SMBH) at its center, the LMC can also help constrain models of early universe BH seed formation as well as the scaling relations of BHs and their host systems in the lower mass range. Here we present a new study of the stellar kinematics near the center of the LMC, and use this to provide the first constraints on the possible presence of a central BH.

\par Over the years, the relations between black-hole mass and properties of their host galaxies such as bulge stellar velocity dispersion, bulge luminosity, and bulge mass have been extensively studied \citep{2000ApJ...539L...9F,2000ApJ...539L..13G,2003ApJ...589L..21M,2004ApJ...604L..89H,2009ApJ...695.1577G,2013ApJ...764..184M,2013ARA&A..51..511K}. While these relations suggest a co-evolution between galaxies and their BHs, they remain poorly constrained for both lower mass black holes (M$_{\rm BH}\lesssim$ 10$^{6}$M$_{\odot}$) and lower mass host systems (M$_{\star}\lesssim$ 10$^{10}$M$_{\odot}$). 
\par In the last 10-15 years, detections of active galactic nuclei (AGN) in nearby dwarf galaxies have provided a means of filling in the lower mass range of the BH mass/host galaxy property relations \citep[e.g.][]{2015ApJ...809L..14B,2011Natur.470...66R,2003ApJ...588L..13F,2004ApJ...607...90B,2015ApJ...809..101D}. Lately, more systematic surveys have been done by sampling larger datasets \citep[e.g.][]{2007ApJ...656...84G,2013ApJ...775..116R}. However the measurements for BHs in dwarf galaxies are still relatively scarce, often have high uncertainties, and none are near enough to study using detailed kinematics of their stars. 
\par Supermassive black holes ($\geq$10$^{6}$M$_{\odot}$) are found in the centers of virtually all massive galaxies \citep[e.g.][]{2013ARA&A..51..511K}. The most distant SMBHs are seen as quasars with redshifts indicating that they existed at a time when the universe was less than a billion years old \citep[e.g.][]{2011Natur.474..616M}. It is still not understood how black holes could become this massive on such a short time scale. In contrast to very massive galaxies, the fraction of dwarf galaxies with massive black holes at their centers is currently unknown. The non-detection of a massive BH in M33 \citep{2001AJ....122.2469G,2001Sci...293.1116M} has shown that the occupation fraction for low-mass galaxies must be lower than unity. A handful of recent studies have placed constraints of M$_{\rm BH}\lesssim$ 10$^{4}$-10$^{6}$M$_{\odot}$ on a few nearby dwarf galaxies \citep[e.g.][]{2011Natur.470...66R,2003ApJ...588L..13F,2009ApJ...690.1031B, 2004ApJ...607...90B, 2005ApJ...628..137V, 2010ApJ...714..713S, 2012AdAst2012E..15N, 2015ApJ...809..101D}. The determination of this fraction for low-mass galaxies (M$_{\star}\lesssim$ 10$^{10}$M$_{\odot}$) can help constrain different theories of the formation of primordial BHs in the first billion years of the universe \citep{2012NatCo...3E1304G}. 
\par One explanation for high-redshift BH formation is the existence of `seed' black holes with M $\lesssim 1000$M$_{\odot}$ produced by the collapse of Population III stars. This requires super-Eddington accretion to explain the rapid growth of SMBHs in the early universe \citep[e.g.][]{2014ApJ...784L..38M}. If this mechanism is the primary source of seed BHs in the early universe, it predicts that nearly all ($>$90\%) present-day low-mass (M$_{\star}\sim10^{9}$M$_{\odot}$) galaxies necessarily contain BHs at their centers \citep{2012NatCo...3E1304G}. On the other hand, another theory proposes the existence of more massive seed black holes of masses on the order of $\sim10^{4-5}$M$_{\odot}$ produced from the direct collapse of pre-galactic disks and gas clouds in the early universe \citep[e.g.][]{2006MNRAS.370..289B,2006MNRAS.371.1813L}. This theory predicts that $\sim$50\% of present-day dwarf galaxies would contain central BHs \citep{2012NatCo...3E1304G}. Depending on the mechanism for these primordial BH's formation, \cite{2009MNRAS.400.1911V} found that the slope and scatter in the $M_{\rm BH}-\sigma$ relation would vary for BH masses $\lesssim10^{6}M_{\odot}$. In either case, it is clear that the search for black holes in the range 10$^{4-6}$M$_{\odot}$ in dwarf galaxies like the LMC can provide insight into the process of black hole formation, growth, and their relationships to their host systems.
\par Alternatively, there may be signatures of IMBHs in nearby globular clusters (GCs). This can be searched for by measuring the velocity dispersion ($\sigma$) of stars near the center. Since GCs around the Milky Way are significantly closer to us relative to almost all other potential hosts of IMBHs, the `sphere-of-influence' of any BH becomes a larger angle on the sky, and therefore possible to probe through stellar motions. The existence of IMBHs in nearby globular clusters remains a topic of ongoing discussion. To date, there has been no evidence of accretion of IMBH's in the centers of GCs through X-ray or radio observations \citep{Maccarone2005,2012ApJ...750L..27S}. Some kinematic evidence for the presence of IMBHs (10$^{4-5}$M$_{\odot}$) in globular clusters has been found from studies with Integral Field Units (IFUs) \citep[e.g.][]{2008ApJ...676.1008N,2011A&A...533A..36L,2013A&A...554A..63F}. However, these claims have been challenged by groups measuring the velocity dispersion with proper motions \citep{2010ApJ...710.1032A,2012ApJ...745..175M} and measurements of individual radial velocities \citep{2013ApJ...769..107L}.  

\begin{figure*}
\plotone{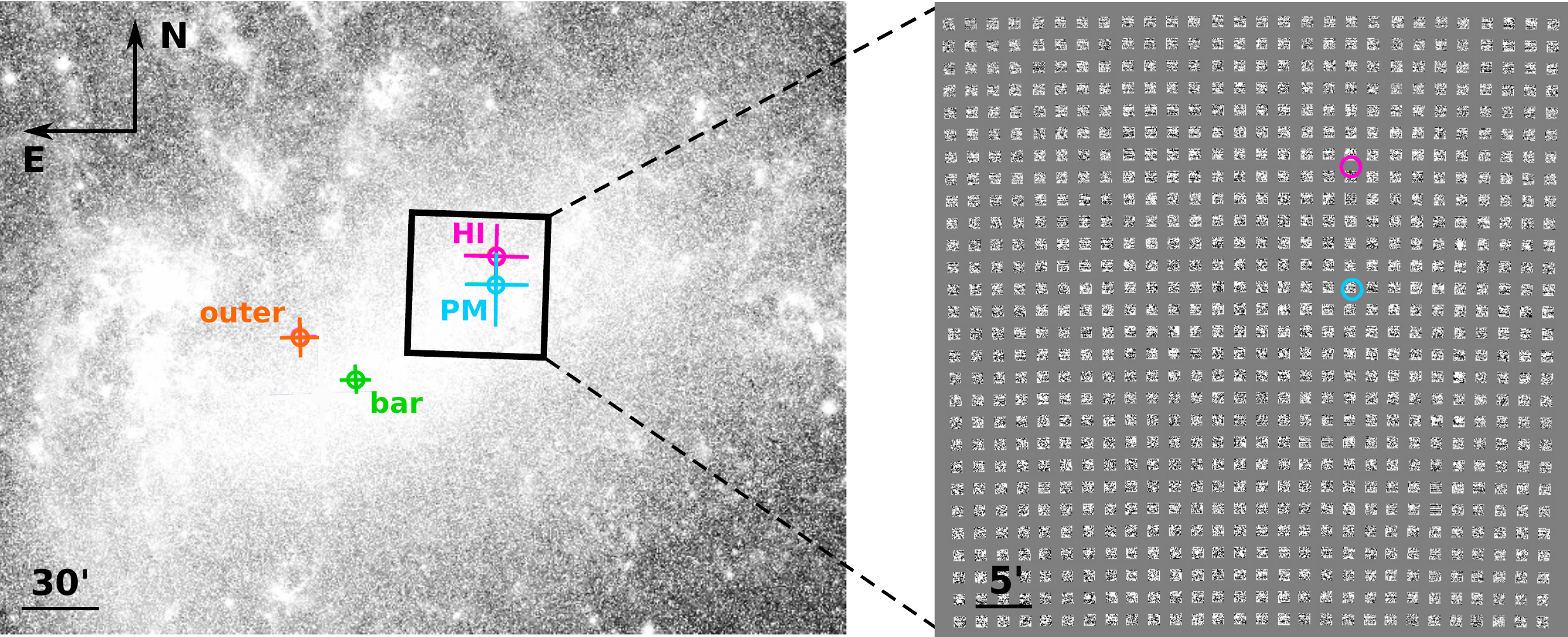}
\caption{The Large Magellanic Cloud with known kinematic and photometric centers. Left panel: central bar of the LMC and its centers determined with different methods. HI: the gas dynamical center of the cold HI disk \citep{1992A&A...263...41L,1998ApJ...503..674K}, PM: the stellar dynamical center inferred from a model fit to proper motions by \cite{2014ApJ...781..121V}, bar: the densest point in the bar \citep{2001AJ....122.1827V,1972VA.....14..163D}, outer: the center of the outer isoplets extracted from the 2MASS survey and corrected for viewing perspective by \cite{2001AJ....122.1827V}. Right panel: The $1\degr\times1\degr$ area observed with MUSE.}
\label{fig:knownCenters}
\end{figure*}

\par Spectroscopic techniques for constraining the presence of BHs in
galaxy centers using stellar or gaseous kinematics are well
established, and typically use a slit or small IFU field placed at the
galaxy center. However, application of these techniques to the LMC
poses the unique challenge that the exact position of its center is
poorly determined. There are two reasons for this. First, since the
LMC is relatively close to us ($\sim50$ kpc away), it spans an
enormous area on the sky. Stars can be traced to $\sim 10^{\circ}$ and
beyond on either side \citep[e.g.][]{2016ApJ...825...20B}. Second, the morphology of the LMC (the prototype of the
class of Magellanic {\it Irregular} galaxies) is asymmetric. The
photometric center differs from the kinematic center by more than a
degree \citep[e.g.][]{2013ApJ...764..161K,2014ApJ...781..121V} (see
Figure \ref{fig:knownCenters}). Also, its kinematics are complex,
disturbed, and poorly understood. These features are due to its
ongoing interaction with the Small Magellanic Cloud (SMC) and Milky
Way
\citep{1993MNRAS.261..873H,2005AJ....129.1465C,2012MNRAS.421.2109B}.

\par Hence, despite many existing studies of the LMC, the position of
its kinematic center is only known to $\sim30$ arcmin. The best
available constraints come from the analysis of the velocity fields of
HI gas \citep{1992A&A...263...41L,1998ApJ...503..674K} and stellar
proper motions \citep{2014ApJ...781..121V}. So to constrain the
possible presence of a central BH, it is necessary to map
spectroscopically an area of about a square degree. This is well
beyond the capabilities of almost all existing spectrographs, given
reasonable amounts of exposure time. However, the most powerful IFU
ever built, the Multi Unit Spectroscopic Explorer (MUSE) instrument,
was recently commissioned on the ESO Very Large Telescope. In this
paper we report the results of using MUSE to map the largest region in
the LMC ever measured spectroscopically in integrated light, using many pointings with
short exposure times over a square degree area surrounding the
kinematic center (see Figure \ref{fig:knownCenters}). Through this
method we combine the velocities of many individual stars to build up
a velocity field and determine the rotation curve, and use these to
search for the kinematic signature of a black hole.

\par Section \ref{MUSE Observations} of the paper details these observations and the data reduction while section \ref{Analysis} describes the construction of the velocity map and subsequent comparison to black hole models. Using this line-of-sight velocity map and the derived rotation curve, we set an upper-mass limit on any black hole within the central degree of the LMC. We put the results into the context of dwarf galaxies and their black holes in section \ref{Discussion}.

\section{MUSE Observations and Data Reduction} \label{MUSE Observations}

Our observations were collected with the Multi Unit Spectroscopic Explorer (MUSE) on the Very Large Telescope (VLT) of ESO’s LA Silla Paranal Observatory in Chile under the programme 094.B-0566. MUSE is a second generation instrument designed for the VLT \citep{2010SPIE.7735E..08B}. It includes an integral field unit (IFU) that operates towards the red in the visual wavelength range (465nm - 930nm). We observed in the wide field mode of the instrument which provided a spatial sampling of 0.2" over a field of view that was 1$\times$1 arcmin$^2$. This mode captures over $90000$ simultaneous spectra in one pointing, and has a resolving power of 2000 at 465nm and 4000 at 930nm. Taking advantage of the excellent spatial coverage provided by this instrument, we observed the central square degree of the LMC with 784 individual pointings separated from one another by 1 arcmin. Therefore the final dataset covers $\sim25\%$ of the central square degree of the LMC. This is illustrated in Figure \ref{fig:knownCenters}.
\par The kinematic centers of the HI and stellar velocity fields and their error bars are contained within this coverage. By running a Monte Carlo simulation of the weighed mean of these two known centers we are 96.5\% certain that the true kinematic center lies within the area covered in our observations. 
Each pointing contains a 309 $\times$ 317 spaxel image provided by the MUSE IFU. This amounts to spectra of thousands of stars in the central region of interest from which we constructed our velocity map.
\par
\begin{figure}
\plotone{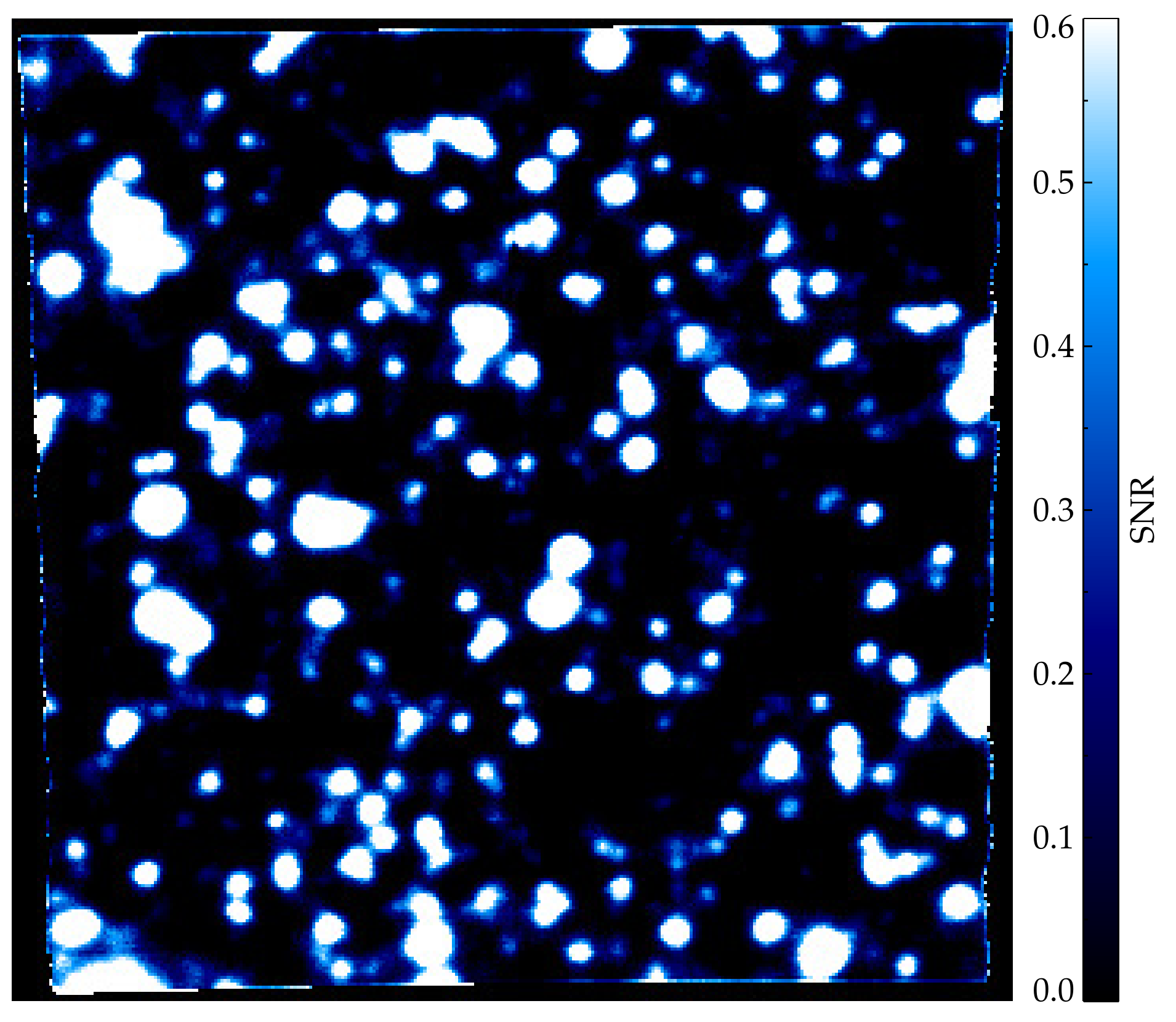}
\caption{Signal-to-noise map of a typical MUSE field. The area of this field is one arcmin$^{2}$.}
\label{fig:snr}
\end{figure}

 Our observations were slated as a filler program on the instrument. Taken on multiple days in November and December 2014, each pointing had an exposure time of only 60 seconds allowing us to map a large area of the sky in a short period of time. By later combining the spectra of all the stars in a single pointing, the signal-to-noise-ratio (SNR) was built up to form integrated spectra from which accurate velocity measurements could be taken. In this way, each pointing in our observations becomes a data point in the velocity map, representing the average velocity for the LMC stars in that field. A representative pointing in SNR units is displayed in Figure \ref{fig:snr}.

\par The data reduction was performed with the ESO pipeline designed specifically for MUSE. The preliminary routines (\texttt{muse\_bias}, \texttt{muse\_dark}, \texttt{muse\_flat}, \texttt{muse\_wavecal}, and \texttt{muse\_skyflat}) combine calibration frames into master calibrations to be used in the subsequent steps (\texttt{muse\_scibasic} and \texttt{muse\_scipost}) which perform the flat fielding, sky subtraction, and wavelength calibration. A detailed documentation of these routines and their functions can be found in the MUSE Pipeline Manual\footnote{\label{note1}Documentation available at https://www.eso.org/sci/software/pipelines/muse/}.

\section{Analysis} \label{Analysis}

Within each MUSE pointing we first identified and excluded foreground sources in the fields before combining all remaining spectra with high enough signal-to-noise into a representative spectrum for that pointing. We then measured a line-of-sight (LOS) velocity from the Calcium Triplet absorption feature ($\sim850$nm) in each spectrum and used this to construct a velocity map. A six-parameter Markov Chain Monte Carlo analysis was used to compare our map to model maps containing a black hole of varying masses. The results set a limit on the mass of any black hole within the center of the LMC. The following subsections detail these steps.

\subsection{Identifying foreground sources} \label{foreground}

To obtain the intrinsic kinematics of the LMC, it was necessary to identify and remove known foreground sources before combining the spectra. To identify all sources within our fields we used the software Source Extractor (SExtractor) \citep{1996A&AS..117..393B} which allowed us to create a catalogue of stars and their positions in each pointing.

\par Largely, the default values of the configuration file were used, with the main alterations tabulated in Table \ref{table:param}. DETECT\_MINAREA is the minimum number of pixels needed to be considered an object. This was increased from a value of 5 to discourage identifying small, bright image artifacts as stellar sources. Assigning the FILTER parameter as N simply turns the keyword off and prevents a process of smoothing the image before detecting pixels. This was done to prevent sources close together from being smoothed into each other. The smoothing process would be helpful with detecting faint extended objects, and is therefore not useful in our crowded star fields. DETECT\_THRESH is the detection threshold for determining objects relative to the background root mean square (RMS) value. Through several trials, a value of 1.5 was determined to best identify the obvious sources in the field. The parameter DEBLEND\_MINCONT controls the program's criteria for determining when bright objects close together are separate sources. The value of 0.001 is the fraction that a number of counts in a separate branch of an object has to be above the total count of the object to be flagged as independent.  Combining the catalogs into a master list of all positions of sources in the data allowed for easy comparison with the 2MASS catalog - where the foreground sources are known.

\begin{table}[h!]
	\centering
	\caption{Altered parameters in the Source Extractor configuration file}
	\label{table:param}
	\begin{tabular}{lc}
		\toprule
		Parameter & Value \\
		\midrule
		DETECT\_MINAREA & 30 \\
		DETECT\_THRESH & 1.5 \\
		FILTER & N \\
		DEBLEND\_MINCONT & 0.001 \\
		\bottomrule
	\end{tabular}
\end{table}

\par Analysis of 2MASS data done by \cite{2000ApJ...542..804N} was used for the identification of foreground stars. The 2MASS survey collected raw imaging data in the near-infrared (NIR) at J (1.25$\mu$m), H (1.65$\mu$m), and K$_{s}$ (2.16$\mu$m) that covered 99.998\% of the celestial sphere between 1997 and 2001 \citep{2006AJ....131.1163S}. The 2MASS All-Sky Point Source Catalog \citep{2006AJ....131.1163S} provided us with the catalog\footnote{Downloaded from http://www.ipac.caltech.edu/2mass/releases/allsky/} of all point source positions and their J, H, and K$_{s}$ magnitudes within a 3500 arcsec square of the LMC central position RA=5h16m57s, DEC=-69d15m35s. To match our catalogue of MUSE sources with the 2MASS source catalogue, we used CataXcorr\footnote{Developed by P. Montegriffo at INAF-Osservatorio Astronomico di Bologna. The package is available at http://davide2.bo.astro.it/~paolo/Main/CataPack.html}, a code specifically developed to perform astrometric matching. This resulted in a complete list of sources in our data for which we had NIR information.

\par
\begin{figure}
\plotone{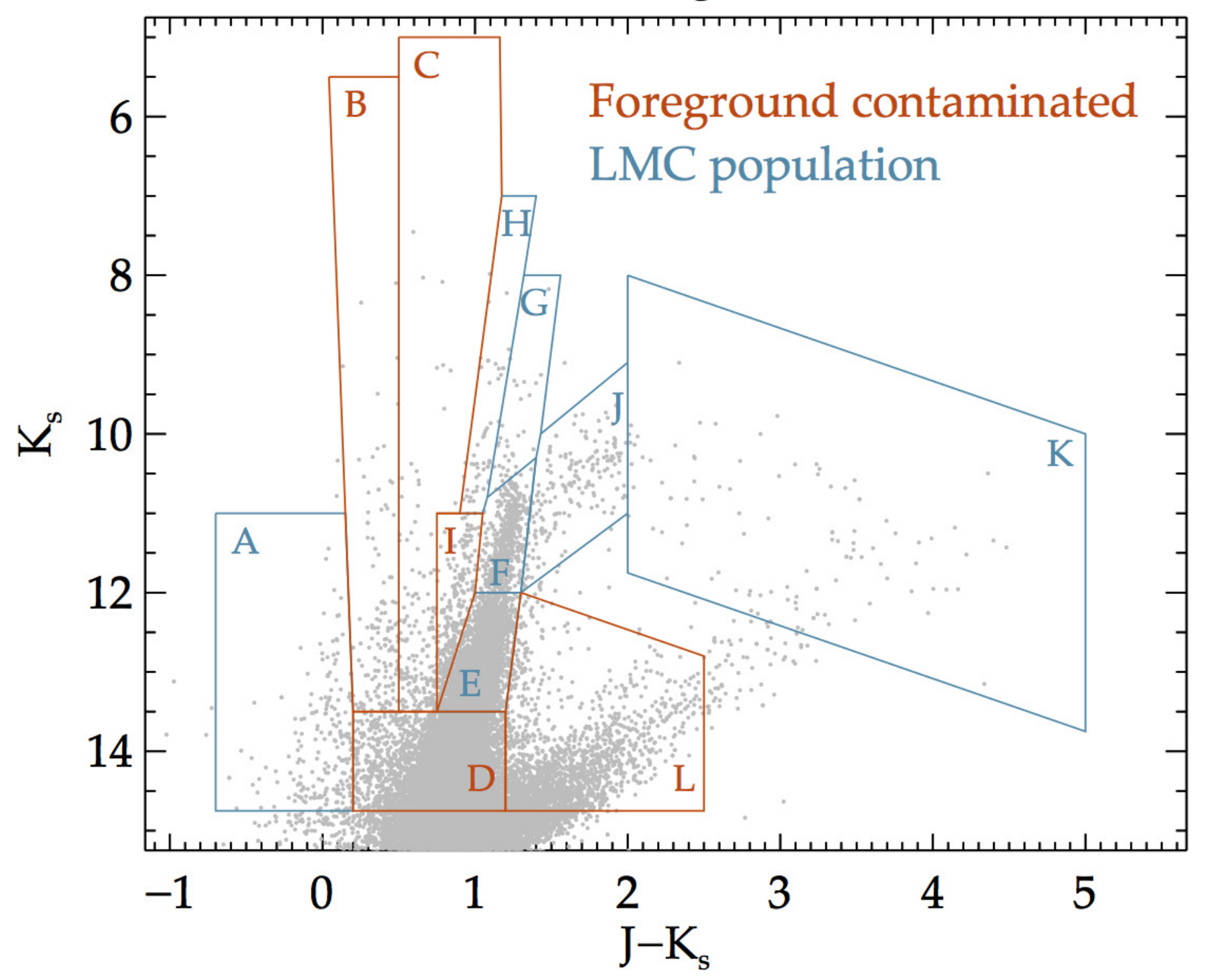}
\caption{Color-magnitude diagram for LMC sources in both 2MASS and MUSE data overlayed with regions as defined by \cite{2000ApJ...542..804N}.}
\label{fig:cmd}
\end{figure}

\cite{2000ApJ...542..804N} identified 12 stellar populations from the color-magnitude diagram (CMD) \citep[Figure 3 of][]{2000ApJ...542..804N} constructed from 2MASS data. Shown in Figure \ref{fig:cmd} is the CMD of the sources identified both in the MUSE data and 2MASS. It is overlayed with the same regions as defined by \cite{2000ApJ...542..804N}.

\par To identify 12 distinct stellar populations, \cite{2000ApJ...542..804N} make use of both the spatial density distribution of 2MASS sources \citep[Figure 4 of][]{2000ApJ...542..804N} and the theoretical colors/isochrones in the CMD. Through visual inspection of the spatial density distribution of the stars, the regions in the CMD that are heavily foreground contaminated are identified as B, C, D, I, and L which correspond to the regions of the same classification on the CMD in Figure \ref{fig:cmd}. Any stars identified to fall within these regions of the CMD are flagged within the MUSE data. For a more detailed analysis of these stellar populations, we refer to \cite{2000ApJ...542..804N}.

\subsection{Extraction of spectra and kinematics} \label{spectra and kinematics}

\begin{figure}
\plotone{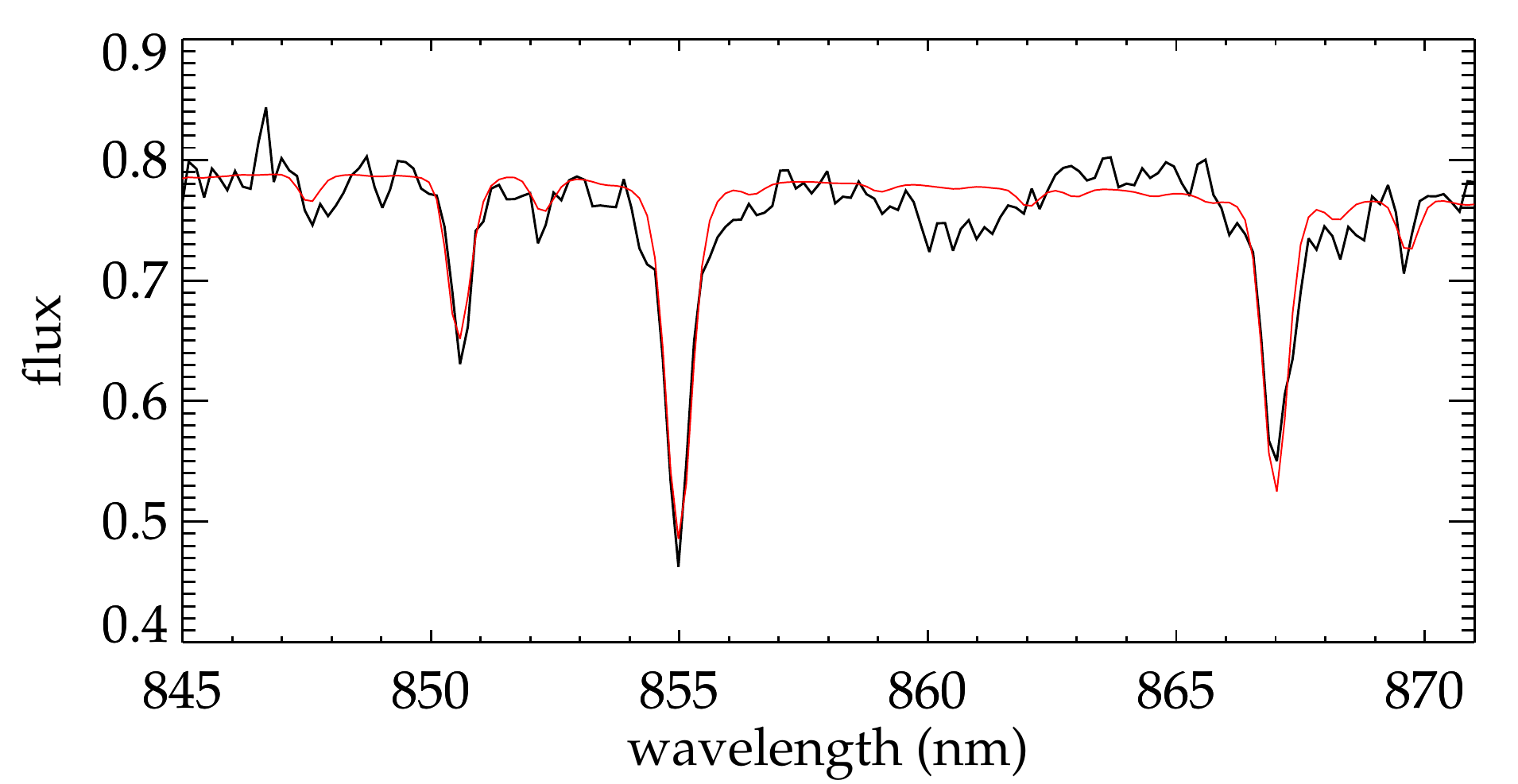}
\caption{Combined spectrum of a single pointing in the region of the Calcium Triplet. Overplotted in red is the bestfit for the spectrum.}
\label{fig:spectrum}
\end{figure}

To create a velocity map, we constructed representative spectra for each of the 784 pointings in our dataset by combining the spaxels we determined to belong to LMC sources. Figure \ref{fig:snr} shows a typical SNR map for a MUSE field in the central region of the LMC. SNR maps like the one displayed are used to identify spaxels that had spectra with SNR $>0.5$ and were therefore  considered to belong to sources in the field. Spaxels belonging to foreground sources identified by the analysis of the infrared CMD in \cite{2000ApJ...542..804N} are excluded in all combined spectra. This was achieved by masking out spaxels within a set radius around the spatial location of each foreground source. The spectra were then combined by computing an iteratively sigma-clipped mean on each pixel in the wavelength dimension. To remove remaining sky signatures not caught in the reduction process, a `sky' spectrum was also generated by combining the SNR $< 0.0$ spaxels. This was then subtracted from the combined signal spectrum to produce the final spectrum of the light from all the LMC stars in the pointing. Each of the 784 spectra representing each MUSE field were made up of an average of 8000 combined spectra within that particular field and had an average SNR of 36.9 in the region of the Calcium Triplet. Figure \ref{fig:spectrum} displays a spectrum representative of the majority of all spectra that were used in measuring radial velocities from the Calcium Triplet absorption lines ($\sim850$ nm). 

\par For each combined spectrum, the penalized pixel-fitting (pPXF) program developed by \cite{2004PASP..116..138C,2016arXiv160708538C} was used to determine both line-of-sight velocity and velocity dispersion. Though the spectral resolution of MUSE is not high enough for a meaningful measurement of the internal LMC velocity dispersion ($\sigma$), an extreme (too high or too low) value of the velocity dispersion inferred from the fit often indicates some problem with the spectra. When fitting models to our kinematic data, the majority of the fields with velocity measurements that were $3\sigma$ away from the best fit also had velocity dispersion measurements greater than $100$ km/s or less than 40 km/s ($\sim$ the instrumental resolution of MUSE). We therefore found these velocity dispersion criteria to be an acceptable method for rejecting spectra with poor velocity measurements. Additionally, $97\%$ of spectra were fit with an average velocity error of 6 km/s. Spectra with velocity errors larger than a 97.5-percentile cut of 26.3 km/s were also considered unreliable.

\par
\begin{figure}
\plotone{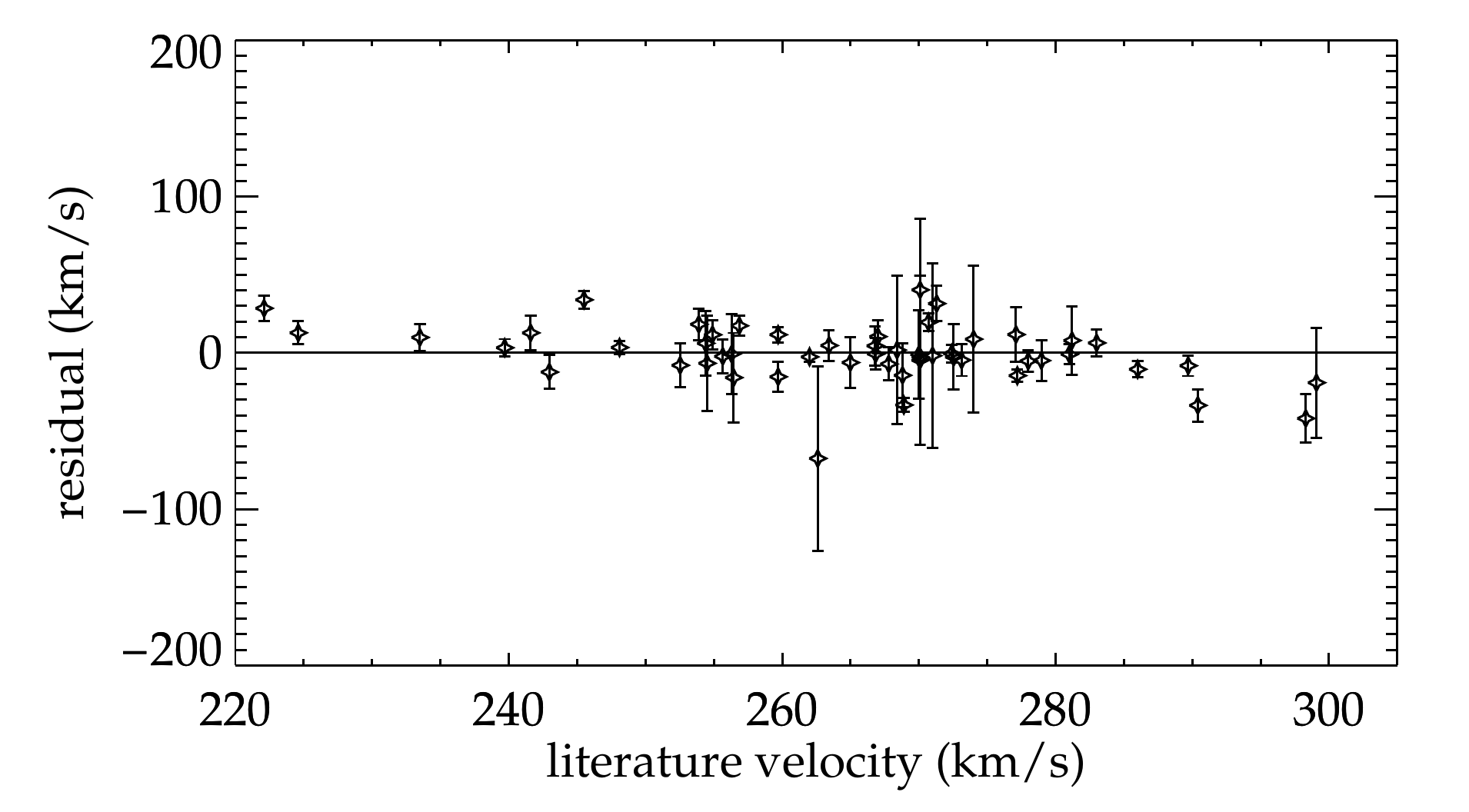}
\caption{Differences between the known velocity of a star in the literature and the measurement we made from the collected MUSE data plotted against the known literature velocities.}
\label{fig:lit_comp}
\end{figure}

 The central degree of the LMC contains 256 stellar sources with known radial velocities in the literature, from the compilation of \cite{2014ApJ...781..121V}. Our $\sim 25\%$ coverage of the central degree of the LMC coincides with 69 of these known LOS measurements. As a consistency check, we derived our own velocities for these sources in our MUSE fields using combined spectra of the spaxels belonging to each of these sources and applying the same pPXF method used on the spectra for our velocity map. On average, the signal-to-noise of the spectra for these individual stars was 8.7 - much less than that of the combined spectra for entire MUSE fields (due to the fact that these spectra were constructed from the combination of considerably fewer spaxels). We reject 11 stars that had velocity dispersion measurements above 100 km/s indicating unreliable velocity measurements, and reject one for having a SNR $< 3$. This left us with 57 acceptable velocity measurements to compare to the known values found in the literature. Figure \ref{fig:lit_comp} displays the differences between our velocity measurements of these stars and the known literature values. Error bars come from the uncertainty in the MUSE measurements where larger error bars correspond to noisier spectra. They are scattered around a weighted mean value of $-0.1 \pm 0.9$ km/s demonstrating that our measurement of individual stellar velocities have a zero-point in agreement with the literature.

\subsection{Modeling} \label{modeling}

\begin{figure*}
\plotone{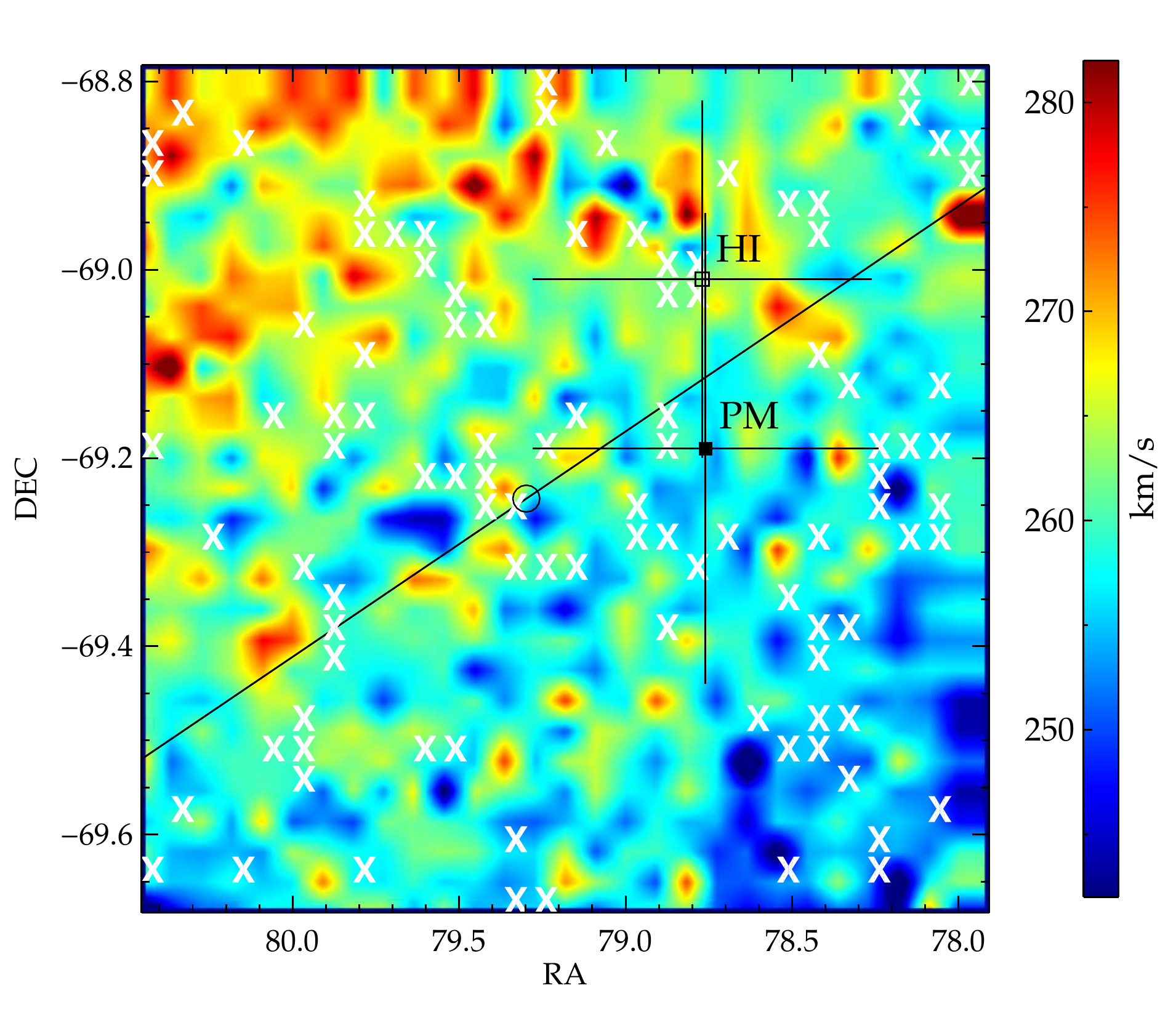}
\caption{The new VLT/MUSE LOS velocity map of the central degree of the LMC. Diagonal line is the central position as fitted by the $x_{0}$ parameter in the five-parameter  model fit. PM is the kinematic center as determined from proper motion data in \cite{2014ApJ...781..121V}, and HI is the HI gas dynamical center as reported in \cite{1992A&A...263...41L,1998ApJ...503..674K}. White `x's indicate a field for which a velocity measurement was excluded on the basis of having $\sigma>$ 100 km/s, $\sigma<$ 40 km/s, $v_{\mathrm{err}}>$ 26 km/s, or lying further than 3-sigma away from the initial model fit. In these locations the color for the plot was filled in with the velocity value from the bestfit model. The open circle is the location where the model fits a black hole with $\sim$1$\sigma$ confidence.}
\label{fig:vmap}
\end{figure*}

Of the 784 velocity points in our map, we discarded 100 of them on the basis of velocity dispersion measurements above 100 km/s, below 40 km/s, or velocity errors above 26 km/s. These criteria were an indication of an unreliable kinematic measurement - see section \ref{spectra and kinematics}. This left us with 684 acceptable points to use as constraints for various models. Figure \ref{fig:vmap} displays our entire 2D velocity field with the rejected points replaced with the bestfit model value at those positions and marked with white crosses. Overplotted with their error bars are the kinematic centers determined in the literature: PM is the stellar kinematic center determined by {\em Hubble Space Telescope} (HST) proper motion measurements \citep{2014ApJ...781..121V}, and HI is the gas dynamical center of the cold HI disk \citep{1992A&A...263...41L,1998ApJ...503..674K}.

\par To characterize our velocity map, we generate models containing two components: a linear velocity field and the velocity field due to the gravitational potential of a BH. Since potentials from different mass components add linearly, and since the circular velocity squared at a given radius is proportional to the radial gradient of the gravitational potential, the circular velocity of our model at a given point \textit{in the plane of the LMC} is given by

\begin{equation}
v_{\rm model} = \sqrt{v_{\rm linear\_lmc}^{2} + v_{\rm BH}^{2}}
\label{eq:model_lmcplane}
\end{equation}
where the circular velocity of the black hole in the inclined LMC plane ($v_{\rm BH}$) is given by

\begin{equation}
v_{\rm BH}^{2} = GM/r
\label{eq:bh}
\end{equation} 
where $M$ is the mass of the black hole in solar masses, $G$ is the gravitational constant in appropriate units, and $r$ is the distance from the black hole in degrees (transformed from kpc using the distance to the LMC as $50.1$ kpc \citep{2001ApJ...553...47F}).

\par $v_{\rm linear\_lmc}$ in equation \ref{eq:model_lmcplane} corresponds to the case of solid-body cylindrical rotation, which is a reasonable approximation for the central regions of of disk galaxies. This linear component was calculated in the plane of the sky as $v_{\rm linear}$ (see equation \ref{eq:plane}) before being transformed into the LMC plane using an inclination angle $34\degr$ \citep{2014ApJ...781..121V}. This linear component is given by

\begin{equation}
v_{\rm linear} = v_{0} + v_{1}*(x'-x_{0})
\label{eq:plane}
\end{equation}
where, $x'$ (and $y'$) are coordinates rotated by an angle $\phi$ up from the horizontal with an origin $(x',y')=(0,0)$ at $(\rm RA,DEC) = (80.45, -69.70)$, $x_{0}$ is the position of the center about which the plane `pivots', $v_{0}$ is set to the systematic velocity of the LMC \citep[261.1 $\pm$ 2.2 km/s][]{2014ApJ...781..121V}, and $v_{1}$ is the slope of the plane in km/s per degree. In these coordinates, the position of a modelled black hole in the plane of the sky is ($x_{0}$, $y_{0}$).

\begin{figure*}
\plotone{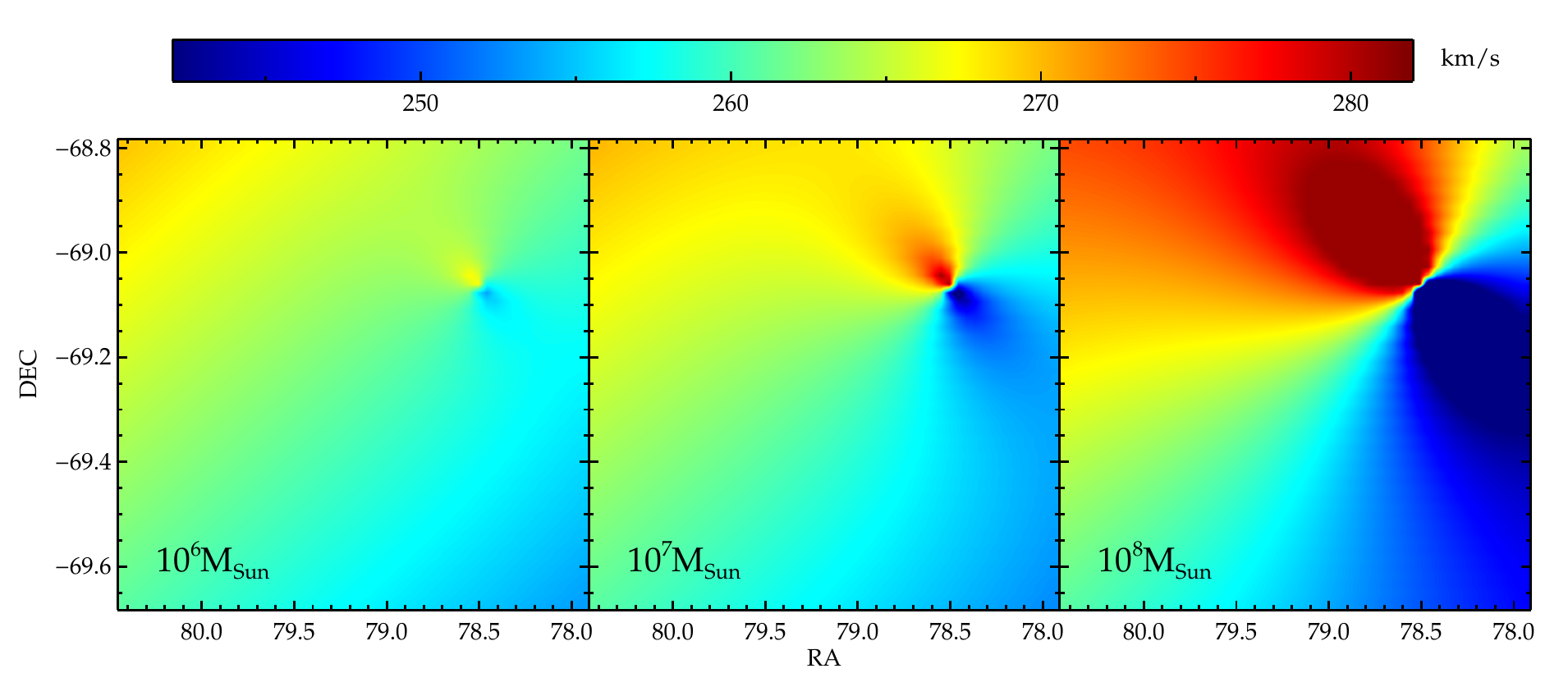}
\caption{Models of the LMC velocity field with $10^{6}$ M$_{\sun}$, $10^{7}$ M$_{\sun}$, and $10^{8}$ M$_{\sun}$ black holes, over the area surrounding the kinematic center implied by previous measurements. In these plots, the black hole was placed at the kinematic center, but its position was varied in our model fits.}
\label{fig:BH_models}
\end{figure*}

The total model velocity ($v_{\rm model}$) at each position was then transformed from the LMC plane into the plane of the sky to produce our model velocity maps. Figure \ref{fig:BH_models} shows example models displaying the signatures of $10^{6}$ M$_{\sun}$, $10^{7}$ M$_{\sun}$, and $10^{8}$ M$_{\sun}$ black holes. Clearly, models containing BHs around $10^{8}$ solar masses are strongly ruled out. To be as accurate as possible in the regions near the black hole, the fields that vary significantly over a single pointing were divided into subgrids. Any field within 0.2 degrees of the black hole position was divided into a subgrid of 50$\times$50 points (avoiding the central singularity) where we evaluated the individual velocity values and then averaged over these to assign a velocity to the entire field. In this way, the model velocity maps in Figure \ref{fig:BH_models} are generated with the same spatial sampling as the map in Figure \ref{fig:vmap}.

\par To efficiently fit our data to this five-parameter model we turn to Markov Chain Monte Carlo (MCMC) analysis. We use the EMCEE package developed by \cite{2013PASP..125..306F}, which is an implementation of the affine-invariant MCMC ensemble sampler by \cite{GoodmanWeare2010}. This was done by defining a log-likelihood quantity:

\begin{equation}
\begin{split}
\ln p(v|x_{0}, & v_{1},\phi,y_{0},\log M_{\mathrm{BH}}) \\ 
 & = -\frac{1}{2}\sum_{n}\Big[\frac{(v_{\mathrm{obs},n}-v_{\mathrm{model},n})^{2}}{s_{n}^{2}}+\ln(2\pi s_{n}^{2})\Big]
\end{split}
\label{eq:lnlike}
\end{equation}
that sums over all valid points in the velocity map. Here, $v_{\rm obs}$ is the measured velocity from the data for that position, $v_{\rm model}$ is the velocity generated by our model, and $s_{n}^{2}$ is given by

\begin{equation}
s_{n}^{2} = v_{\mathrm{err},n}^{2}+f^{2}(v_{\mathrm{model},n})^{2}
\label{eq:sn}
\end{equation}
where $v_{\rm err}$ is the error obtained by running a Monte Carlo routine with the pPXF software \citep{2004PASP..116..138C,2016arXiv160708538C}, and $f$ is a sixth parameter of the MCMC that we allowed to vary as the fractional amount our errors could be underestimated by.

\par After running the MCMC over the 684 points in our map, we rejected an additional field that had a velocity value more than 3-sigma away from the best-fit model: leaving us with 683 fields. This field was determined to be contaminated by an unusually bright foreground star.

\begin{figure*}
\plotone{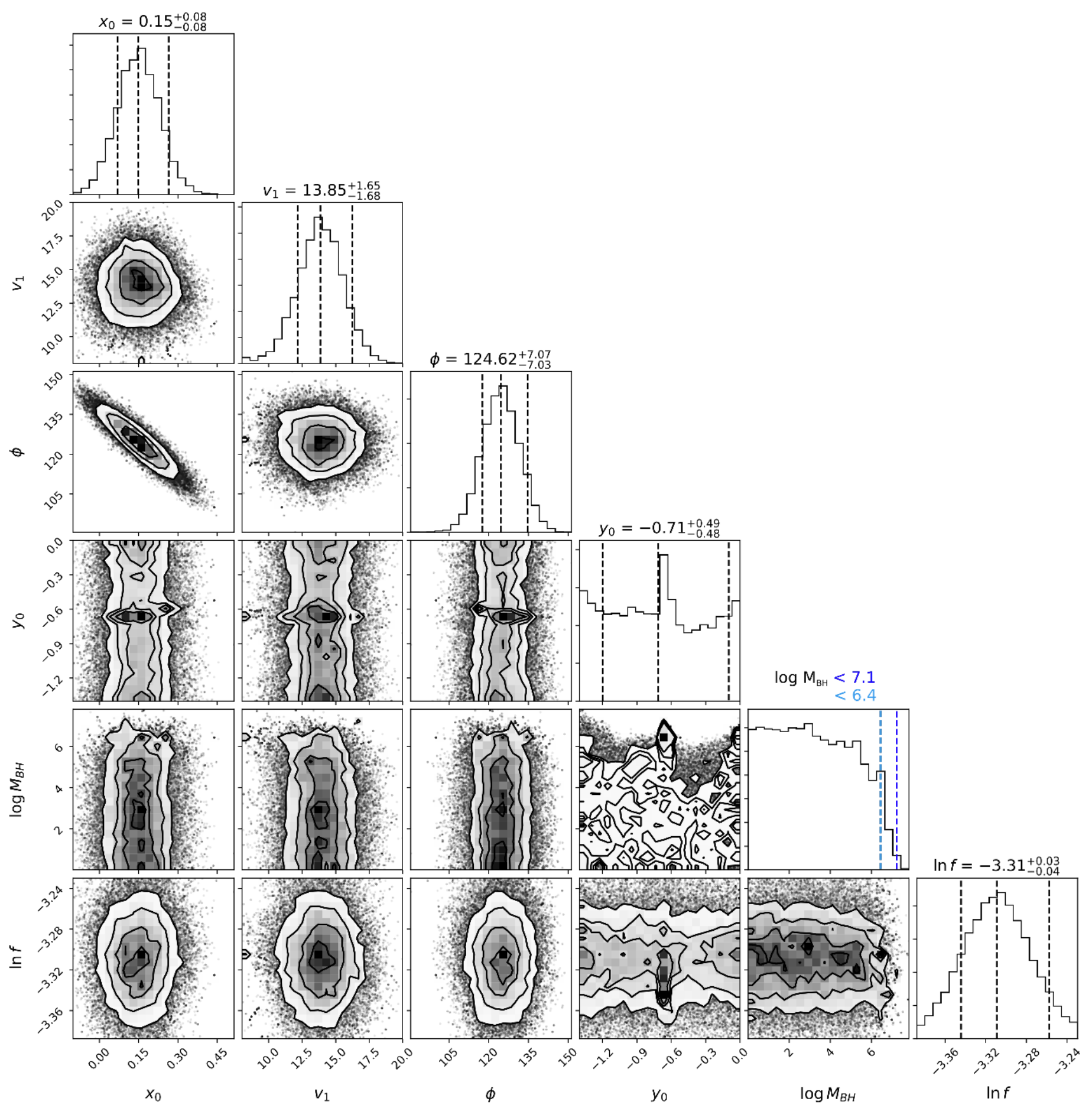}
\caption{MCMC distributions for the velocity model fit to our data. Scatter plots are the projected two dimensional distributions of the posterior probabilities of the parameters. Histograms display the one dimensional distributions with vertical dashed lines drawn at the 16th, 50th, and 84th percentiles. The $\log M_{BH}$ histogram instead plots dark and light blue lines at the 99.7th (3$\sigma$) and 95.5th (2$\sigma$) percentiles respectively. From top-to-bottom and left-to-right, the panels display: $x_{0}$ (the position in degrees of the `central' velocity line at 261 km/s), $v_{1}$ (the linear component of velocity model in km/s per degree), $\phi$ (the angle that the normal to the $x_{0}$ axis makes with the horizontal), $y_{0}$ (the position of the black hole), $\log M_{BH}$ (the logarithm of the mass of the black hole in solar masses), and $\ln f$ (the natural logarithm of the fractional amount the errors are underestimated.) }
\label{fig:cornerplot}
\end{figure*}

\par All five parameters of our velocity model plus $\ln f$ are varied in the MCMC and compared with the 683 reliable measurements in the velocity field. Figure \ref{fig:cornerplot} shows the result of the MCMC with the bestfit parameters as $x_{0}=0.15 \pm 0.08$ degrees, $v_{1}=13.85^{+1.65}_{-1.68}$ km/s per degree, $\phi=124.62^{+7.07}_{-7.03}$ degrees, $y_{0}=-0.71^{+0.49}_{-0.48}$ degrees, and $\ln f=-3.31^{+0.03}_{-0.04}$, with 1$\sigma$ confidence. The significantly small value of $\ln f$ is an indication that our errors ($v_{\rm err}$) are not underestimated. Plotted in the histogram for $\log M_{\rm BH}$ in Figure \ref{fig:cornerplot} are 99.7th (3$\sigma$) and 95.5th (2$\sigma$) percentiles, indicating the upper limits on the mass for any black hole within the central degree of the LMC to be $10^{7.1}$ M$_{\sun}$ and $10^{6.4}$ M$_{\sun}$ respectively. We found the observable signature of black holes of masses around $10^{7}$ M$_{\sun}$ to be slightly stronger than the natural fluctuation noise within our map. Our models containing a black hole of a mass in the range $10^{5-6}$ M$_{\sun}$ fit the data with $1\sigma$ confidence at a spatial position of $y_{0}$ = -0.65 degrees. This is plotted as an open circle on the data in Figure \ref{fig:vmap}. However, considering that the detection has low statistical significance, and that the signatures of these sizes of black holes closely resemble the fluctuations in our data, we do not interpret this fit to indicate the presence of a black hole.

\par In a separate analysis to the MCMC fit, we also constrained the presence of a black hole through a $\chi^{2}$ minimization. By first fitting a 2D plane described by equation \ref{eq:plane}, we fit for the parameters $x_{0}$, $v_{1}$, and $\phi$ simultaneously. Then holding these three constant at their bestfit values, we performed the $\chi^{2}$ minimization using equation \ref{eq:model_lmcplane} as our model, varying both the position ($y_{0}$) and mass (M$_{\rm BH}$) of the black hole. The results yielded values consistent with the MCMC analysis. Most notably, the limit on the mass of the black hole did not change.

\subsection{One dimensional rotation curve}
\begin{figure*}
\plotone{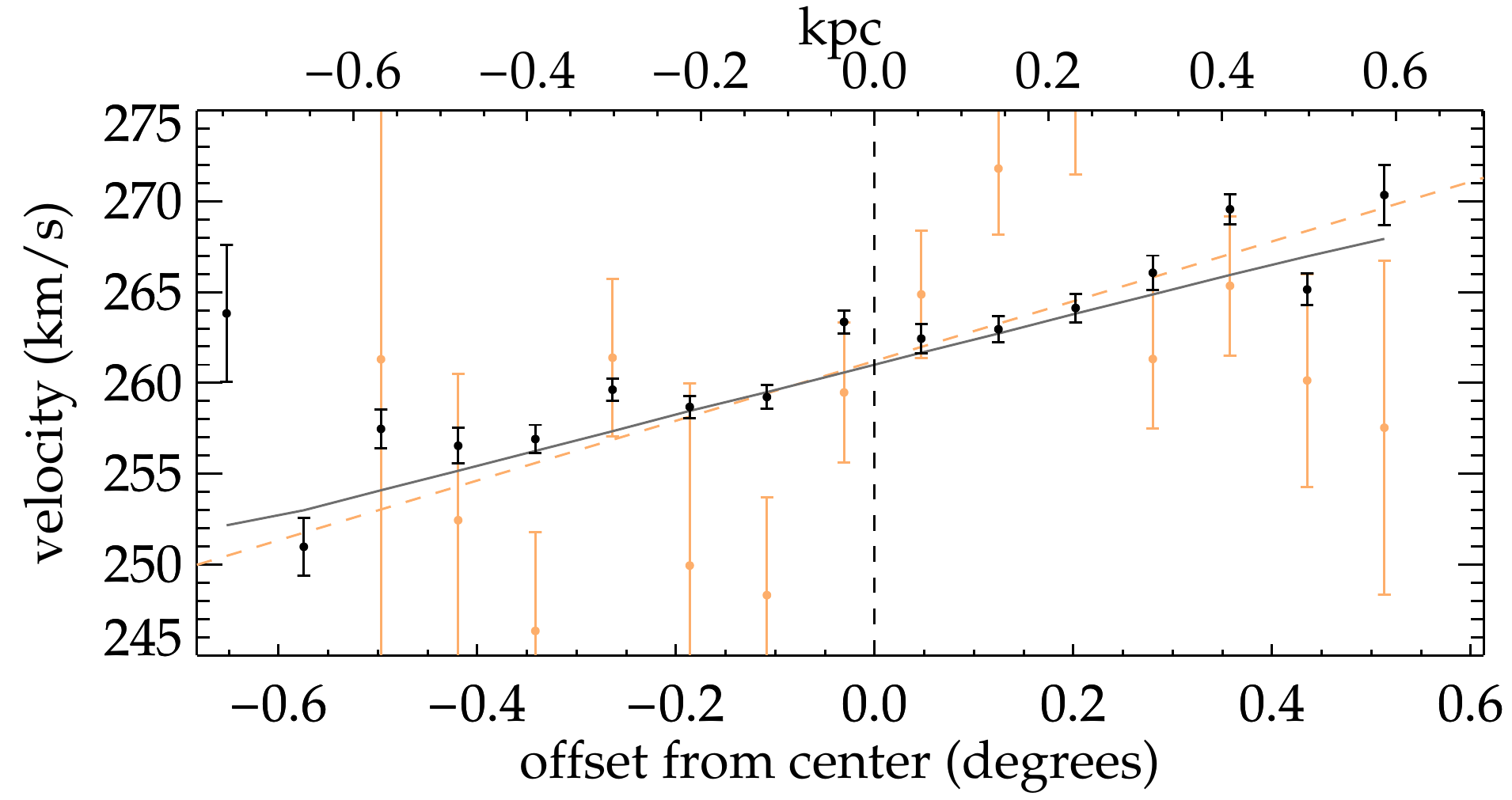}
\caption{One dimensional rotation curve for the LMC averaged over the entire field. Each black data point was found by taking a weighted mean of the velocity values from the 2D map in bins along the central line fit. Overplotted in orange are the binned LOS velocities for 256 individual stars compiled from the literature in \cite{2014ApJ...781..121V} in our square degree field of view. The dashed line shows the best linear fit to these data.}
\label{fig:1D_noslice}
\end{figure*}

\begin{figure*}
\plotone{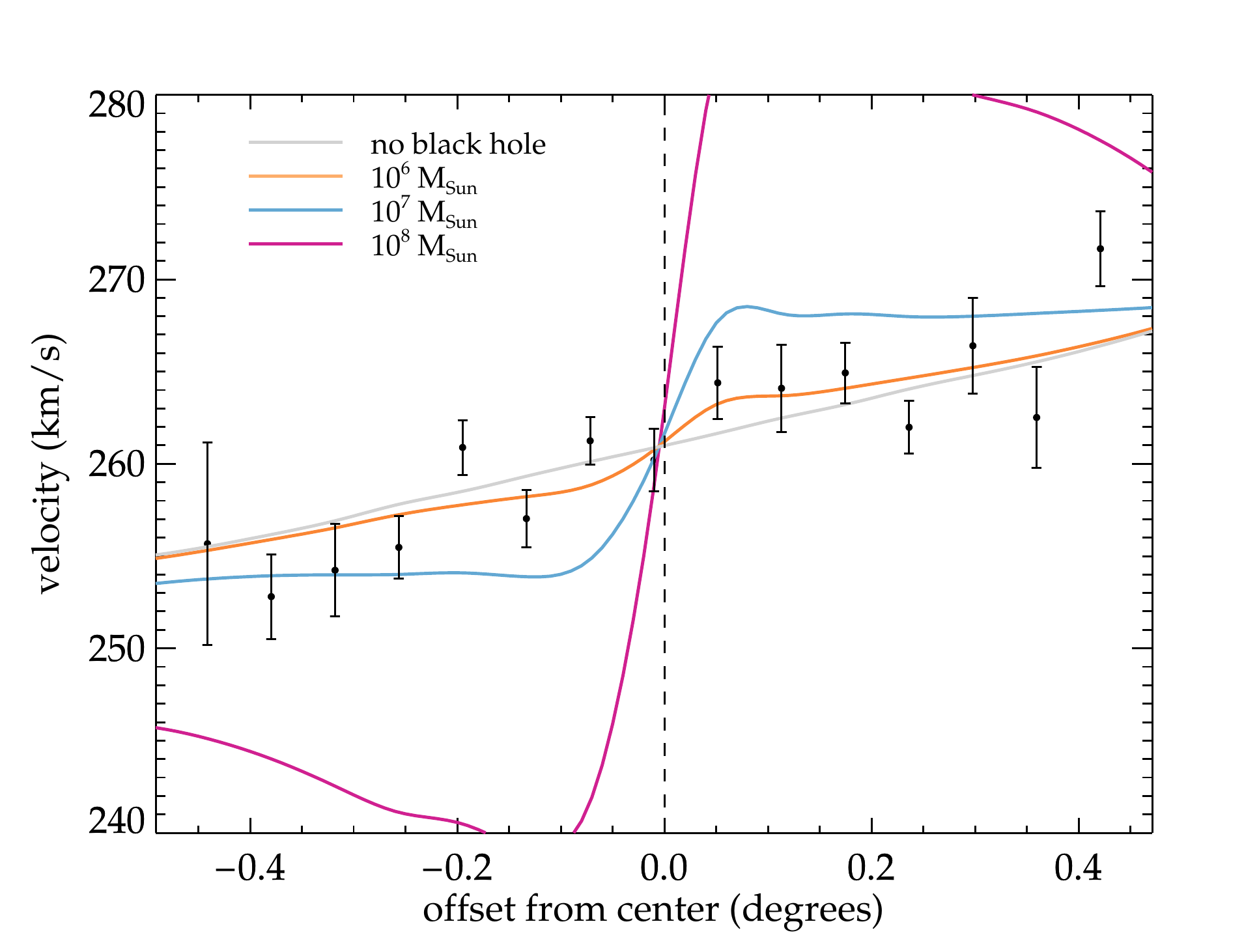}
\caption{One dimensional rotation curve for the LMC at the average position of the known kinematic center. Overplotted are the model rotation curves for fields with black holes of masses $10^{6}$ M$_{\sun}$ , $10^{7}$ M$_{\sun}$ , and $10^{8}$ M$_{\sun}$ at this location. Each data point in the one dimensional curve was found by taking a weighted mean of velocity values from the 2D map within a 0.3 degree slice of the kinematic center.}
\label{fig:1D_BHs}
\end{figure*}

The one dimensional rotation curve of the data was produced by collapsing the velocity map along the axis perpendicular to the `central' line fit to the two dimensional plane. This is displayed in Figure \ref{fig:1D_noslice}. Alternatively, Figure \ref{fig:1D_BHs} displays the rotation curve in the region of the kinematic center with the rotation curves for black holes of varying masses overplotted. We performed a $\chi^{2}$-fit on this rotation curve of our data with the rotation curves of black hole models, which yielded results consistent with the 2D fit.  

\par Considering the entire central degree, the best known LOS model of this area comes from analysis done by \cite{2014ApJ...781..121V} using global LOS velocities from various sources in the literature combined with large scale \textit{HST} proper motions in the LMC. Binning LOS data from the 256 individual stars in the study within our coverage area (displayed as the orange points in Figure \ref{fig:1D_noslice}) results in a slope of the rotation curve of 15.3 $\pm$ 5.8 km/s per degree. This is consistent with the slope of 13.9 $\pm$ 1.6 km/s per degree measured from our MUSE data and illustrates the significant improvement that our measurement for the rotation curve has over the previous data available.

\section{Discussion and Conclusions} \label{Discussion}

The center of the Large Magellanic Cloud is an enticing place to look for a central black hole. If it hosts a BH, the proximity to our own galaxy would mean that this dwarf-galaxy/low-mass BH system is readily available to study in extreme detail. This could be a very significant test of the $M_{\rm BH}-\sigma$ relation between BHs and their host galaxies at the lower mass end and would contribute to constraining models of SMBH growth early in the universe \citep{2012NatCo...3E1304G}.

\par We have presented here the most detailed velocity field measurement for the center of the LMC to date, based on measurements with VLT/MUSE. We have used these new data to constrain the possible presence of a central BH in the LMC. We arrive at a 3$\sigma$ upper-mass limit of $10^{7.1}$ M$_{\sun}$ for a black hole at the center of the LMC, or a 2$\sigma$ upper limit of $10^{6.4}$ M$_{\sun}$. We also report the slope of the rotation curve over the central region of the LMC to be measured with an improved precision of 13.9 $\pm$ 1.6 km/s per degree (or 15.8 $\pm$ 1.8 km/s per kpc). We found this measurement to be in agreement with the slope derived from binned measurements of LOS velocities from individual stars in the literature, but with a factor of four smaller uncertainty.
\par
\begin{figure*}
\plotone{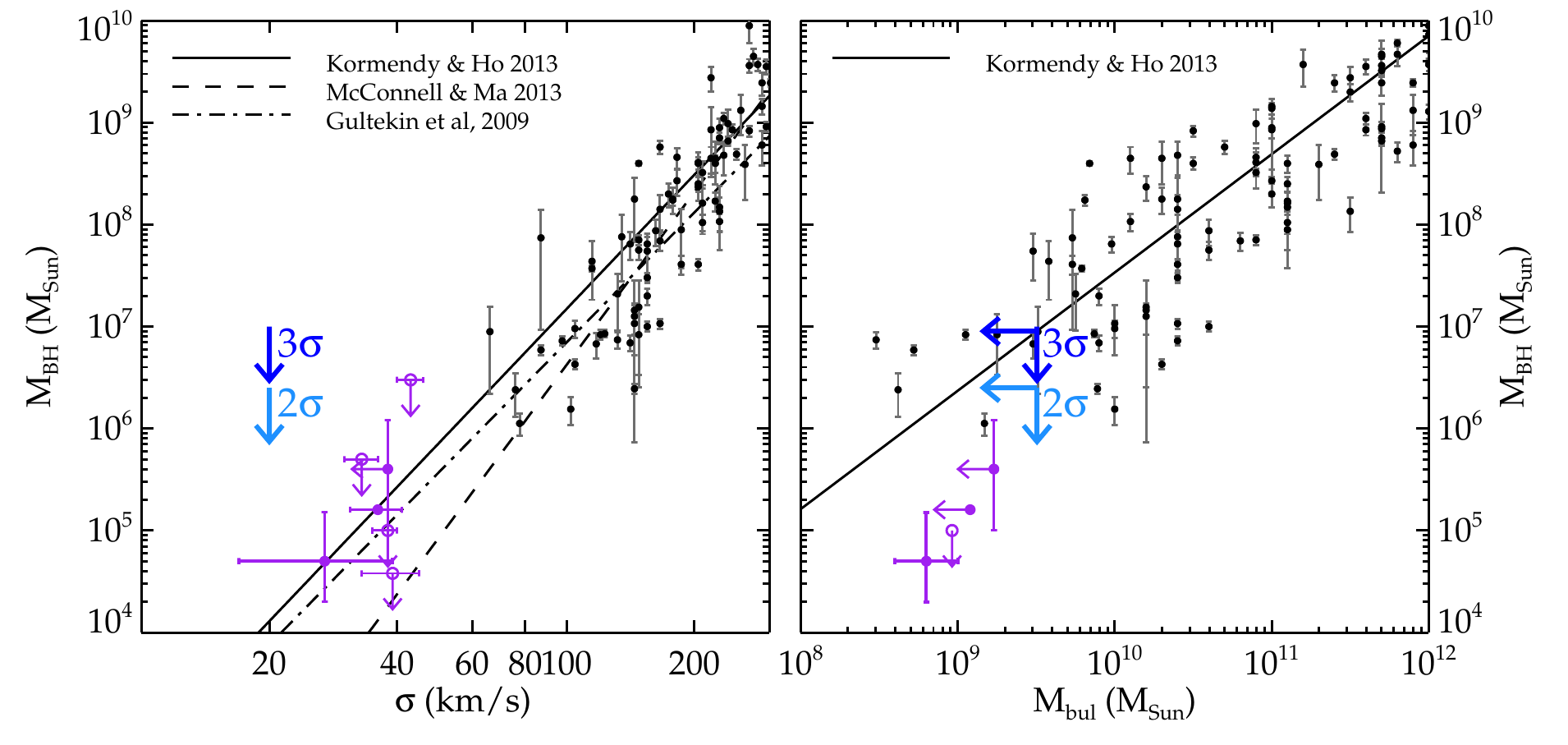}
\caption{The M$_{\rm BH}$-$\sigma$ (left) and the M$_{\rm BH}$-M$_{\rm bul}$ (right) correlation for galaxies hosting supermassive black holes. Black data points are detections of BHs that come from the recent survey and compilation \cite{2016ApJ...818...47S}. Black solid and dashed lines represent various determinations of the scaling relations in other works \citep{2013ARA&A..51..511K,2013ApJ...764..184M,2009ApJ...695.1577G}. Our 3$\sigma$ and 2$\sigma$ upper limits for the BH mass in the LMC are indicated by the blue arrows. For the LMC we use the disk velocity dispersion of $\sigma_{\rm disk}\sim20$ km/s \citep{2014ApJ...781..121V} and use the total baryonic mass of the LMC as an upper limit at M$_{\rm gal}\sim3.2\times10^{9}$M$_{\odot}$ \citep{2002AJ....124.2639V} since the LMC is a disk galaxy without a classical bulge (so $M_{\rm bul}\ll M_{\rm gal}$). Also plotted in filled purple circles are detections of black holes at the low mass end: the smallest BH ever reported in a galactic nucleus RGG 118 \citep{2015ApJ...809L..14B}, the well studied dwarf system POX 52 \citep{2004ApJ...607...90B, 2008ApJ...686..892T}, and the small and bulgeless galaxy NGC 4395 \citep{2003ApJ...588L..13F,2005ApJ...632..799P,2015ApJ...809..101D}. Open circles are upper-mass limits for BHs in low-$\sigma$ galaxies: NGC 404 \citep{2010ApJ...714..713S}, IC 342 \citep{1999AJ....118..831B}, NGC 205 \citep{2005ApJ...628..137V}, and NGC 3621 \citep{2009ApJ...690.1031B}. }
\label{fig:imbh}
\end{figure*}

 Shown in Figure \ref{fig:imbh} are two of the scaling relations observed between supermassive black holes and their host systems. The left panel describes the relationship between black-hole mass and the velocity dispersion of its host system, while the right describes the relationship between the black-hole mass and its host system's bulge-mass. Extrapolating these relations to a lower mass range leads to the consideration of dwarf galaxies, nuclear clusters, and globular clusters with intermediate-mass black holes of masses $<10^{6}$M$_{\odot}$. Our upper limit for the mass of a black hole in the LMC is plotted in blue at both 2$\sigma$ and 3$\sigma$ confidence. They are plotted against estimates for the disk velocity dispersion \citep[$\sigma_{\rm disk}\sim20$ km/s][]{2014ApJ...781..121V} and the total baryonic mass of the LMC \citep[M$_{\rm gal}\sim3.2\times10^{9}$M$_{\odot}$][]{2002AJ....124.2639V}. It should be noted though that the known BH scaling relations for other galaxies pertain to galaxy bulges. The LMC is a late-type Magellanic irregular disk galaxy. It does not have a well-defined bulge, and hence $M_{\rm bul}\ll M_{\rm gal}$. Moreover, while $\sigma_{\rm disk}$ for the LMC is well-measured, it is not clear whether it is meaningful to interpret it in the same way as for galaxy bulges. One alternative is to use the quadratic sum of the disk velocity dispersion and the rotational velocity of the LMC: $\sigma=\sqrt{\sigma_{\rm disk}^{2}+v_{\rm rot}^{2}}\sim93$ km/s, where $v_{\rm rot}=91.7\pm18.8$ km/s \citep{2014ApJ...781..121V}. This would shift our upper limits to the right on the M-sigma relation. With these caveats in mind, our results are not inconsistent with any of the known scaling relations. 

\par Also shown in Figure \ref{fig:imbh} are three notable measurements of IMBHs found in other dwarf galaxies in the last 15 years: POX 52 \citep{2004ApJ...607...90B, 2008ApJ...686..892T},  NGC 4395 \citep{2003ApJ...588L..13F,2005ApJ...632..799P,2015ApJ...809..101D}, and RGG 118 \citep{2015ApJ...809L..14B}. Upper limits from other dynamical methods are also plotted for the nearby S0 type galaxy NGC 404 \citep{2010ApJ...714..713S}, the nuclear star cluster of spiral galaxy IC 342 \citep{1999AJ....118..831B}, the dwarf elliptical galaxy NGC 205 \citep{2005ApJ...628..137V}, and the bulgeless spiral galaxy NGC 3621 \citep{2009ApJ...690.1031B}. Due to the difficulty in detecting black holes in this mass range, and the distance to these objects limiting our ability to measure detailed kinematics, measurements in this regime remain scarce. 

\par The limit we place on a BH in the LMC is unique in mapping individual stellar motions within the central 0.5 kpc of the host galaxy. This is the first measured upper-mass limit for a central black hole in the Large Magellanic Cloud and by far the nearest dwarf galaxy with a kinematic limit on a central black hole mass. The limit re-enforces the expectation that if the LMC harbors any massive black hole, it is in the intermediate-mass range. This new result is also of interest in the context of the detection of candidate hypervelocity stars in/from the LMC \citep[e.g.][]{2016arXiv161105504L,2008ApJ...675L..77B,2008ApJ...684L.103P,2016ApJ...825L...6B}, which could be connected to the possible presence of a central black hole in the LMC.
\par This study is a promising step in understanding the complex dynamics of the LMC's central bar region. Performing a similar analysis of the central region over a smaller area with higher resolution and longer integration time could easily constrain this upper-limit further. Or, if the LMC does harbor an IMBH, velocity maps generated from integral field spectrographs such as MUSE could very well detect its kinematic signature. Additionally, using our spectra to identify distinct stellar populations and generate velocity maps of these different populations would provide further insight into the complex dynamics and history of the LMC's bar. 
\par
Since the relations between black hole mass and their host systems hold for galaxies with a wide range of properties, the growth of a central black hole seems to be closely linked with the process of galaxy formation. The study of intermediate-mass black holes in dwarf galaxies is therefore crucial for the understanding black-hole growth and interaction with its surrounding host system. It is still unknown what fraction of low-mass galaxies contain black holes, and the measurements remain difficult to achieve. By studying one of the closest low-mass galaxies, the LMC, we can get a relatively close up view of a dwarf galaxy and what the effects that any black hole has on its components and structure. Further kinematic study could unveil any black hole lurking within our complicated neighbor, especially if the position of its kinematic center can be pinned down more accurately.

\acknowledgments
This work was supported in part by the Leiden/ESA Astrophysics Program for Summer Students (LEAPS) and the Space Telescope Science Institute's (STScI) Space Astronomy Summer Program. Based on observations collected at the European Organisation for Astronomical Research in the Southern Hemisphere, Chile under ESO programme 094.B-0566. This publication makes use of data products from the Two Micron All Sky Survey, which is a joint project of the University of Massachusetts and the Infrared Processing and Analysis Center/California Institute of Technology, funded by the National Aeronautics and Space Administration and the National Science Foundation. We would like to thank the anonymous referee for helpful comments and suggestions.

\bibliographystyle{aasjournal}
\bibliography{boyce_LMC_revised}{}

\end{document}